# Implications for the Cosmological Landscape: Can Thermal Inputs from a Prior Universe Account for Relic Graviton Production?

A.W. Beckwith[1]

[1] *APS member, Menlo Park, CA .*
abeckwith@UH.edu

**Abstract.** Sean Carroll's pre-inflation state of low temperature-low entropy provides a bridge between two models with different predictions. The Wheeler-de Witt equation provides thermal input into today's universe for graviton production. Also, brane world models by Sundrum allow low entropy conditions, as given by Carroll & Chen (2005). Moreover, this paper answers the question of how to go from a brane world model to the 10 to the 32 power Kelvin conditions stated by Weinberg in 1972 as necessary for the initiation of quantum gravity processes. This is a way of getting around the fact CMBR is cut off at a red shift of z = 1100. This paper discusses the difference in values of the upper bound of the cosmological constant between a large upper bound predicated for a temperature dependent vacuum energy predicted by Park (2002), and the much lower bound predicted by Barvinsky (2006). with the difference in values in vacuum energy contributing to relic graviton production. This paper claims that this large thermal influx, with a high initial cosmological constant and a large region of space for relic gravitons interacting with space-time up to the z = 1100 CMBR observational limit are interlinked processes delineated in the Lloyd (2002) analogy of the universe as a quantum computing system. Finally, the paper claims that linking a shrinking prior universe via a worm hole solution for a pseudo time dependent Wheeler-De Witt equation permits graviton generation as thermal input from the prior universe, transferred instantaneously to relic inflationary conditions today. The existence of a wormhole is presented as a necessary condition for relic gravitons. Proving the sufficiency of the existence of a worm hole for relic gravitons is a future project.

## INTRODUCTION

First, it is necessary to consider if there is an inherent fluctuation in early universe cosmology linked to a vacuum state nucleating out of "nothing." Such a vacuum state has been found, and its nucleation leads to relic graviton production. The vacuum fluctuation leads to production of a dark energy density. This is initially due to contributions from an axion wall, which is dissolved during the inflationary era. What we will be doing is to reconcile how that wall was dissolved in early universe cosmology with quantum gravity models, brane world models.Weinberg (1972) predicting of a threshold of 10 to the 32 power Kelvin for quantum effects becomes dominant in quantum gravity models. These findings suggest that relic graviton production could account for the presence of strong gravitational fields at the onset inflation, as described by Guth (1981,2003), which would be in line with the predictions by Penrose (1989), based on the Jeans inequality .

It is noteworthy that Barvinsky (2006) recently predicted a range of four-dimensional Planck's constant values between upper and lower bounds. This suggests that the existence of a cosmological constant at about a Planck's time $t_P$ is consistent with the formation of scale factors which permit the existence of definable space-time metrics.

An argument can be made that prior to Planck's time $t_P$, conventional space-time metrics, even those adapting to strongly curved space, do not apply. Park (2003) predicted an upper range of cosmological constant values far greater than that of Barvinsky (2006). The difference can be explained in terms of a thermal/vacuum-energy input into graviton production. To do this, this paper first examines how the Friedman equation provides an evolution of the scale factor $a(t)$ in two cases: (1) with a constant cosmological constant, and (2) when the cosmological constant is far larger than it is today.

Making the case for an initially very large cosmological constant will lead to a roadmap for solving the landscape problem. I.e., $\Lambda_{\max}(Barvinsky) = 3m_P^2/2B = 360m_P^2$ as a peak value after graviton production would lead to a Hartle-Hawking's universe wave function, if we have G = gravitational constant= $6.67428 \times 10^{-11}$ (meter)$^3$kg$^{-1}$/(sec)$^{2:}$

$$\psi_{HH}\big|_{Barvinsky} \approx \exp(-S_E) = \exp(3 \cdot \pi/2 \cdot G\Lambda_{4-Dim}) \neq 0 \tag{1}$$

Whereas, the Park (2002) value for a nearly infinite cosmological vacuum parameter (due to high temperatures) would lead to (prior to graviton production) using the five dimensional, negative cosmological vacuum energy as an embedding structure. This is for a five-dimensional brane world structure containing the four-dimensional high-temperature Park (2002) value for an almost infinite four dimensional cosmological vacuum energy. These results would lead to the following behavior of the Hartle Hawking wave functional as identified by Barvinsky (2006) and modified by Beckwith (2007):

$$\psi_{HH}\big|_{Brane-world} \approx \exp(-S_E) = \exp(3 \cdot \pi/2 \cdot G\Lambda_{5-Dim}) \xrightarrow[T \to \infty]{} 0 \tag{2}$$

This is in part due to the assumption that the absolute value of the five-dimensional "vacuum state" parameter varies with temperature T, as Beckwith (2007) writes

$$\big|\Lambda_{5-\dim}\big| \approx c_1 \cdot \left(1/T^\alpha\right) \tag{3}$$

in contrast with the more traditional four dimensional version, without the minus sign of the brane world theory version. The five=dimensional version is actually connected with Brane theory and higher dimensions, whereas the four dimensional version is linked to more traditional De Sitter space-time geometry, as given by Park (2002)

$$\Lambda_{4-\dim} \approx c_2 \cdot T^\beta \tag{4}$$

This would allow for making inroads toward a solution to the cosmological land scape problem discussed by Guth (2003). I.e, why have $10^{1000}$ or so independent vacuum states as predicted by string theory? To do this, this paper starts with a maximum entropy formulation of the initial expansion of the universe by reconciling what happens in conventional inflationary cosmology. A zero valued scalar field due to initial exponential expansion of the scale factor, which becomes extremely complicated up to where the scale factor is proportional to time to the 2/3rds power.. But one can improve this analysis by arguing for conditions that give a short term quintessence contribution to dark energy, an outgrowth of the quantum computer analogy given below.

## PRELIMINARY ANALOGY USING SETH LLOYD'S PAPER ON THE UNIVERSE AS A QUANTUM COMPUTER

The formula given by Lloyd (2002) regarding the number of operations the "Universe" can "compute" during its evolution, if $\hbar \approx 1.05457 \times 10^{-34} \; kg \cdot (meter)^2/\sec$, is:

$$\#operations = \frac{4E}{\hbar} \cdot \left(t_1 - \sqrt{t_1 t_0}\right) \approx \left(t_{Final}/t_P\right) \leq 10^{120} \tag{5}$$

Assume that $t_1$ = final time of physical evolution, where $t_0 = t_P \sim 10^{-43}$ seconds, and that we can set an energy input by assuming, in early-universe conditions, that a variable parameter $N^+$ which is $\neq$ parameter $\varepsilon^+$ where $0 \leq \varepsilon^+ \ll 1$. So we write an energy value as

$$E = (V_{4-Dim}) \cdot \left[ \rho_{Vac} = \frac{\Lambda}{8\pi G} \right] \sim N^+ \cdot \left[ \rho_{graviton} \cdot V_{4-vol} \approx \hbar \cdot \omega_{graviton} \right] \quad (6)$$

This assumes $\rho_{graviton} \sim 8.2089591(81) \times 10^{95}$ kg/(meter)$^3$, as well as for a four dimensional volume $V_{4-vol} \sim$ three dimensional volume, plus time in the way Carroll (2004) describes via $d^4x = dx^0 \Lambda_w ... \Lambda_w dx^3$, with the wedge products $\Lambda_w$ defined, as an example, for two 1 forms via $(A\Lambda_w B)_{u,v} = A_u B_v - A_v B_u$, and our formula for incremental four-space "volume" is a product of typical three space geometry, with time multiplied in. Furthermore, if we assume the temperature is close to $T \approx 10^{29}$ Kelvin initially, if

$$H = \sqrt{8\pi G \cdot [\rho_{crit} \sim \rho_{graviton}]/3 \cdot c^2} \text{ and } \rho_{crit} \sim \rho_{graviton} \sim \hbar \cdot \omega_{graviton}/V_{4-Vol}, \quad (7)$$

Then,

$$\# operations \approx 1/[t_P^2 \cdot H] \approx \sqrt{V_{4-Vol}} \cdot t_P^{-2} / \sqrt{[8\pi G \hbar \omega_{graviton}/3c^2]} \approx [3\ln 2/4]^{4/3} \cdot [S_{Entrophy}/k_B \ln 2]^{4/3} \quad (8)$$

So, using the principle of maximum entropy, this leads to a long-term time duration of relic graviton interaction with space-time up to the z = 1100 red shift CMBR barrier, with the gravitons initially created with regards to the cosmological constant which abruptly changes with two different values, one given by Park et al, effectively nearly infinite in value, and a subsequent far lower value given by Barvinsky at or about 10 to the minus 43 seconds.

## STATEMENT OF THE GENERAL PROBLEM WE ARE INVESTIGATING

If one looks at allowed upper bounds of the cosmological constant, the difference between what Barvinsky (2006) recently predicted and Park's (2002) upper limit (based on thermal input) shows that a phase transition is occurring at or before Planck's time. It is assumed that a release in gravitons occurs, leading to a removal of graviton energy stored contributions to this cosmological parameter, with $m_P$ as the Planck mass, i.e., the mass of a black hole of "radius" on the order of magnitude of Planck length $l_P \sim 10^{-35}$ m. This leads to Planck's mass $m_P \approx 2.17645 \times 10^{-8}$ kilograms

$$\Lambda_{4-dim} \propto c_2 \cdot T \xrightarrow{graviton-production} 360 \cdot m_P^2 \ll c_2 \cdot [T \approx 10^{32} K] \quad (9)$$

Right after the gravitons are released, there is a drop-off of temperature contributions to the cosmological constant. So for small time values $t \approx \delta^1 \cdot t_P$, $0 < \delta^1 \leq 1$ and for temperatures sharply lower than $T \approx 10^{12} Kelvin$, Beckwith (2006,2007), for a positive integer $n$,

$$\frac{\Lambda_{4-\dim}}{|\Lambda_{5-\dim}|} - 1 \approx \frac{1}{n} \tag{10}$$

This can be done for length ( radii ) values proportional to the inverse of the Hubble parameter when

$$L \propto L(\Lambda_{4-\dim} \approx |\Lambda_{5-\dim}|) \propto H^{-1}(\Lambda_{4-\dim} \approx |\Lambda_{5-\dim}|) \tag{11}$$

Initially, temperatures are on the order of 1.4 times 10 to the 32 power Kelvin as a threshhold for the existence of quantum effects. This would pick the value given by Weinberg (1972) for quantum effects to happen,

$$E_{critical} \equiv 1.22 \times 10^{28} \, eV \tag{12}$$

This assumes a working cosmology that gets to such temperatures at the instance of quantum nucleation of a new universe. And if there is no temperature dependence, in the 5$^{th}$ dimensional cosmological constant, one gets a five-dimensional line element, as given by Wesson (1999), where the cosmological parameter, $\Lambda$, is a constant:

$$dS^2_{5-\dim} \cong \frac{\Lambda \cdot l^2}{3} \cdot \{4-\dim \, Schwartzshield \, deSitter \, metric\} - dl^2 \tag{13}$$

**TABLE 1.** With respect to phenomenology.

| Time | Thermal inputs | Dynamics of axion | Graviton Eqn. |
|---|---|---|---|
| Time $0 \leq t \ll t_P$ | Use of quantum gravity to give thermal input via quantum bounce from prior universe collapse to singularity. Brane theory predicts beginning of graviton production. | Axion wall dominant feature of pre inflation conditions, due to Jeans inequality with enhanced gravitational field, Quintessence scalar equation of motion valid for short time interval | Wheeler formula for relic graviton production beginning to produce gravitons due to sharp rise in temperatures. |
| Time $0 \leq t < t_P$ | End of thermal input from quantum gravity due to prior universe quantum bounce. Brane theory predicts massive relic graviton production | Axion wall is in process of disappearing due to mark rise in temperatures. Quintessence valid for short time interval | Wheeler formula for relic graviton production produces massive spike gravitons due to sharp rise in temperatures |
| Time $0 < t \approx t_P$ | Relic graviton production largely tapering off, due to thermal input rising above a preferred level, via brane theory calculations. Beginning of regime where the $\Lambda_{4-Dim}$ is associated with Guth style inflation. | Axion wall disappears, and beginning of Guth style inflation. Quintessence scalar equations are valid. Beginning of regime for $\frac{\Lambda_{4-\dim}}{|\Lambda_{5-\dim}|} - 1 \approx \frac{1}{n}$ 5 dim $\rightarrow$ 4 dim | Wheeler formula for relic graviton production leading to few relic gravitons being produced. |
| Time $t > t_P$ | No relic graviton production. Brane theory use scalar potential as given by Sago, et. al. | Approaching regime for which super nova survey called Essence applies. 4 dim only | Wheeler formula for relic graviton production gives no gravitons. |

Also, one can expect a difference in the upper limit of Park's four dimensional inflation value for high temperatures, on the order of 10 to the 32 Kelvin, and the upper bound, as Barvinsky (2006) predicts. If put into the Harkle-Hawking's wave function, this diffenence is equivalent to a nucleation-quantization condition, which, it is claimed, is a way to delineate a solution to the cosmic landscape problem that Guth (1981,2000,2003) discussed. In order to reference this argument, it is useful to note that Barvinsky ( 2006) came up with

$$\Lambda_{\max}\big|_{Barvinsky} \cong 360 \cdot m_P^2 \tag{14}$$

A minimum value of

$$\Lambda_{\min}\big|_{Barvinsky} \cong 8.99 \cdot m_P^2 \tag{15}$$

This is in contrast to the nearly infinite value of the Planck's constant as given by Park (2002)

$\cdot \Lambda_{4-dim}$ is defined by Park (2002).with $\varepsilon* = \dfrac{U_T^4}{k^*}$ and $U_T \propto (external\ temperature)$, and $k^* = \left(\dfrac{1}{'AdS\ curvature}\right)$ so that

$$\Lambda_{4-dim,Max}\big|_{Park} \xrightarrow[T \mapsto 10^{32}\ Kelvin]{} \infty \tag{16}$$

As opposed to a minimum value as given by Park (2002)

$$\cdot \Lambda_{4-dim} = 8 \cdot M_5^3 \cdot k^* \cdot \varepsilon^* \xrightarrow[external\ temperature \to 3\ Kelvin]{} (.0004eV)^4 \tag{17}$$

**TABLE 2**.What can be said about cosmological $\Lambda$ in 5 and 4 dimensions.

| **Time** $0 \leq t \ll t_P$ | **Time** $0 \leq t < t_P$ | **Time** $0 < t \approx t_P$ | **Time** $t > t_P \to$ **today** |
|---|---|---|---|
| $\|\Lambda_5\|$ undefined, $T \approx \varepsilon^+ \to T \approx 10^{32} K$ $\Lambda_{4-dim} \approx$ almost $\infty$ | $\|\Lambda_5\| \approx \varepsilon^+$, $\Lambda_{4-dim} \approx$ extremely large $T \approx 10^{12} K$ | $\|\Lambda_5\| \approx \Lambda_{4-dim}$, $T$ smaller than $T \approx 10^{12} K$ | $\|\Lambda_5\| \approx$ huge, $\Lambda_{4-dim} \approx$ small, $T \approx 3.2 K$ |

## WHEELER GRAVITON PRODUCTION FORMULA FOR RELIC GRAVITONS

As is well known, a good statement about the number of gravitons per unit volume with frequencies between $\omega$ and $\omega + d\omega$ may be given by (assuming here, that $\bar{k} = 1.38 \times 10^{-16} erg/^0K$, and $^0K$ is denoting Kelvin temperatures, where Gravitons have two independent polarization states), as given by Weinberg (1972).

$$n(\omega)d\omega = \frac{\omega^2 d\omega}{\pi^2} \cdot \left[\exp\left(\frac{2\cdot\pi\cdot\hbar\cdot\omega}{\bar{k}T}\right) - 1\right]^{-1} \tag{18}$$

The hypothesis presented here is that input thermal energy (given by the prior universe) inputted into an initial cavity/region (dominated by an initially configured low temperature axion domain wall) would be thermally excited to reach the regime of temperature excitation. This would permit an order-of-magnitude drop of axion density $\rho_a$ from an initial temperature $T_{dS}\big|_{t \leq t_P} \sim H_0 \approx 10^{-33} eV$.

## GRAVITON POWER BURST/ WHERE DID THE MISSING CONTRIBUTIONS TO THE COSMOLOGICAL 'CONSTANT' PARAMETER GO?

To do this, one needs to refer to a power spectrum value that can be associated with the emission of a graviton. Fortunately, the literature contains a working expression of power generation for a graviton produced for a rod spinning at a frequency per second $\omega$, per Fontana (2005), for a rod of length $\hat{L}$ and of mass m a formula for graviton production power. This is a variant of a formula given by Park (1955), with mass $m_{graviton} \propto 10^{-60} kg$

$$P(power) = 2 \cdot \frac{m_{graviton}^2 \cdot \hat{L}^4 \cdot \omega_{net}^6}{45 \cdot (c^5 \cdot G)} \tag{19}$$

The contribution of frequency here needs to be understood as a mechanical analogue to the brute mechanics of graviton production. The frequency $\omega_{net}$ is set as an input from an energy value, with graviton production number (in terms of energy) derived via an integration of Eqn. (19) above, $\hat{L} \propto l_P$. This value assumes a huge number of relic gravitons are being produced, due to the temperature variation.

$$\langle n(\omega) \rangle = \frac{1}{\omega(net\ value)} \int_{\omega 1}^{\omega 2} \frac{\omega^2 d\omega}{\pi^2} \cdot \left[\exp\left(\frac{2\cdot\pi\cdot\hbar\cdot\omega}{\bar{k}T}\right) - 1\right]^{-1} \tag{20}$$

And then one can set a normalized "energy input " as $E_{eff} \equiv \langle n(\omega) \rangle \cdot \omega \equiv \omega_{eff}$; with $\hbar\omega \xrightarrow{\hbar \equiv 1} \omega \equiv |E_{critical}|$ given in Eqn. (12) above, which leads to the following table of results, where $T^*$ is an initial temperature of the pre-inflationary universe condition.

**TABLE 3**. Graviton burst.

| Numerical values of graviton production | Scaled Power values |
|---|---|
| N1= $1.794 \times 10^{-6}$ for $Temp = T^*$ | Power = 0 |
| N2= $1.133 \times 10^{-4}$ for $Temp = 2T^*$ | Power = 0 |
| N3= $7.872 \times 10^{+21}$ for $Temp = 3T^*$ | Power = $1.058 \times 10^{+16}$ |
| N4= $3.612 \times 10^{+16}$ for $Temp = 4T^*$ | Power $\cong$ very small value |
| N5= $4.205 \times 10^{-3}$ for $Temp = 5T^*$ | Power= 0 |

Here, N1 refers to a net graviton numerical production value as given by Eqn. (20). There is a distinct power spike that is congruent with a relic graviton burst. The next task is to understand how this graviton burst would be linked to the evolution of a scale factor. This is a datum that is claimed in Eqn. (42) below, which argues for a causal discontinuity between a prior universe giving thermal input for the expressed graviton burst above, and how the scale factor evolves in later times.

## RELIC GRAVITON LINKAGE TO WORMHOLE SOLUTION OF WHEELER-DE WITT EQUATION AND ITS LINKAGE TO A PRIOR UNIVERSE

This paper claims that a prior universe is linked, via a wormhole solution of the Wheeler-De Witt equation, to a present universe configuration, due to a pseudo time component to the Wheeler-De Witt equation. . This wormhole solution is a necessary and sufficient condition for thermal transfer of heat from that prior universe, to allow for graviton production under relic inflationary conditions. To model this, results from Crowell (2005) on quantum fluctuations in space-time are used. This gives a model from a pseudo time component version of the Wheeler-De Witt equation, with use of the Reinssner-Nordstrom metric to help obtain a solution that passes through a thin shell separating two space-times. The radius of the shell $r_0(t)$ separating the two space-times is of length $l_P$ in approximate magnitude, leading to the denominator of the time component for the Reissner-Nordstrom metric

$$dS^2 = -F(r) \cdot dt^2 + \frac{dr^2}{F(r)} + d\Omega^2 \qquad (21)$$

This has:

$$F(r) = 1 - \frac{2M}{r} + \frac{Q^2}{r^2} - \frac{\Lambda}{3} \cdot r^2 \xrightarrow[T \to 10^{32} Kelvin \sim \infty]{} -\frac{\Lambda}{3} \cdot (r = l_P)^2 \qquad (22)$$

This assumes that the cosmological vacuum energy parameter has a temperature dependence as outlined by Park (2002). If

$$\frac{\partial F}{\partial r} \sim -2 \cdot \frac{\Lambda}{3} \cdot (r \approx l_P) \equiv \eta(T) \cdot (r \approx l_P) \qquad (23)$$

then this leads to a wave functional solution to a Wheeler-De Witt equation bridging two space-times. This solution is similar to that being made between these two space-times with "instantaneous" transfer of thermal heat ,as given by Crowell (2005)

$$\Psi(T) \propto -A \cdot \{\eta^2 \cdot C_1\} + A \cdot \eta \cdot \omega^2 \cdot C_2 \qquad (24)$$

This has $C_1 = C_1(\omega,t,r)$ as a pseudo cyclic and evolving function in terms of frequency, time, and spatial function, with the same thing describable about $C_2 = C_2(\omega,t,r)$ with $C_1 = C_1(\omega,t,r) \neq C_2(\omega,t,r)$. The upshot of this is that a thermal bridge between a shrinking prior universe, collapsing to a singularity, and an expanding universe expanding from a singularity exits. Also, an almost instantaneous transfer of heat with terms dominated by $\eta(T)$ exits, and is forming a necessary and sufficient condition for the thermal heat flux.

# EVOLUTION OF THE SCALE FACTOR USING PADAMANBHAN'S LINKAGE BETWEEN DEFINED SCALE FACTORS, POTENTIALS, AND SCALAR FIELDS

Padmanabhan (2006) has described how to form scalar potentials, scalar fields and how this ties in with inflationary cosmological considerations in calculating the red shift. The implications are quite startling. If the scale factor, $a(t)$ is known, is suggests, as Padmanabhan (2006) wrote for the potential and the scalar field,

$$V(t) \equiv V(\phi) \sim \frac{3H^2}{8\pi G} \cdot \left(1 + \frac{\dot{H}}{3H^2}\right), \tag{25}$$

$$\phi(t) \sim \int dt \cdot \sqrt{\frac{-\dot{H}}{4\pi G}} \tag{26}$$

Case I: $a(t) \sim t^\lambda$

Referring to the scale factor and the potential,

$$\phi \sim \sqrt{\frac{\lambda}{4\pi G}} \ln t, \tag{27}$$

$$V(\phi) \sim \frac{3}{8\pi G} \cdot \left(\lambda^2 - \frac{\lambda}{3}\right) \cdot \exp\left[-\sqrt{\frac{4\pi G}{\lambda}} \cdot \phi\right] \tag{28}$$

Case IA: $a(t) \sim t^\lambda : \lambda \geq \frac{2}{3}$,

$$V(\phi) \sim \frac{6}{72\pi G} \cdot \exp\left[-\sqrt{6\pi G} \cdot \phi\right] \approx \frac{6}{72\pi G} \cdot \left[1 - \sqrt{6\pi G} \cdot \phi + 3\pi G \cdot \phi^2 - ..\right] \tag{29}$$

The above approaches a constant value if $\phi \sim \sqrt{\frac{2}{3\pi G}}$, assuming that G is in this case comparatively large a contribution and the denominator is a much larger contribution than the numerator. So in this case

$$V(\phi) \xrightarrow[\phi \to \sqrt{\frac{2}{3\pi G}}]{} \frac{6}{72\pi G} \tag{30}$$

Case II: $a(t) \sim \exp(\tilde{\beta} \cdot t)$

The exponential expansion of the scale factor is now examined. The potential becomes a constant, and the scalar field disappears:

$$\phi \sim 0^+, \tag{31}$$

$$V \sim \frac{3\tilde{\beta}^2}{8\pi G} = \text{constant value} \tag{32}$$

Case III: $a(t) \sim \exp\left(\tilde{\tilde{\beta}} \cdot t^f\right) : 0 < f < 1$

$$\phi \sim \sqrt{-\frac{\tilde{\tilde{\beta}} \cdot f \cdot (f-1)}{8\pi G \cdot (f-2)}} \cdot t^{\left(\frac{f-1}{2}\right)} \tag{33}$$

And,

$$V \sim \left[\frac{3\tilde{\tilde{\beta}}^2 f^2}{8\pi G}\right] \cdot \left(B_1 \phi^{C1} + B_2 \phi^{C2}\right) \tag{34}$$

Where: $C1 = (4f - 4/f - 1)$, $C2 = (2f - 4/f - 1)$, $B_1 = \left[-\frac{8\pi G \cdot (f-2) \cdot (f-1)}{f \cdot \tilde{\tilde{\beta}}}\right]^{\frac{2f-2}{f-1}}$,

$$B_2 = \left[-\frac{8\pi G \cdot (f-2) \cdot (f-1)}{f \cdot \tilde{\tilde{\beta}}}\right]^{\frac{f-2}{f-1}} \cdot \left[\frac{f \cdot (f-1)}{3f^2}\right]$$

Case IV: As given by Padmanabhan

$$V \sim \lambda \phi^4 \tag{35}$$

This will lead to the following scale factor and scalar field with respect to time values:

$$\phi \sim \phi_i \cdot \exp\left[-\sqrt{\frac{32\lambda \cdot M_P^2}{6}} \cdot (t - t_i)\right] \tag{36}$$

And,

$$a(t) \sim a_i \cdot \exp\left(\frac{\phi_i^2}{8M_P^2} \cdot \left[1 - \exp\left[-\sqrt{\frac{64\lambda M_P^2}{3}} \cdot (t - t_i)\right]\right]\right) \tag{37}$$

All these four cases are applicable where there is a continuum of values between the immediate aftermath of the big bang, up to inflation, and to models purporting to show a speedup of expansion due to a small cosmological constant, as will be discussed in this text.

The upshot is that graviton production is required as a way of linking the symmetry that exists between Case II (which in the beginning of cosmological evolution has a zero scalar field that pops up) and its reappearance in a resumption of an expanding, accelerating universe due to a reduced cosmological constant value. That assumes a scalar field that would be now evolving, after the $a(t) \sim t^{2/3}$ matter-dominated era, as:

$$a(t) \sim \left[\sinh\left(\frac{3}{2} \cdot \sqrt{\frac{\Lambda_{3-\text{deg},Kelvin}}{3}} \cdot ct\right)\right]^{2/3} \tag{38}$$

Eventually, this tends toward an exponential expansion rate of the scale factor, of the form given below

$$a(t) \sim \left[\exp\left(\frac{3}{2} \cdot \sqrt{\frac{\Lambda_{3-deg, Kelvin}}{3}} \cdot ct\right)\right] \quad (39)$$

This would lead to, once again, a zero quintessence field and a constant potential, as in CASE II above.

For short time intervals, close to the point when the scalar field is first nucleated, we go to a nonlinear regime where one is solving a Friedman equation, stated by Frampton (2007), where $\rho_{rel}$, an $\rho_{matter}$ are given as density values in a cyclic universe treatment of the Friedman equation given by Frampton (2007)

$$(\dot{a}/a)^2 = \frac{8\pi G}{3} \cdot [\rho_{rel} + \rho_{matter}] + \frac{\Lambda}{3} \quad (40)$$

For small scale-factor values initial values of the relativistic and matter density inputs into the Friedman equation we denote by subscripts, small o, as for the scale factor and the two types of matter and radiation energy density factors. This leads to, as we argue below,

$$\int da \frac{1}{\left[a^4(t) + \frac{8\pi}{\Lambda}\left[(\rho_{rel})_0 a_0^4 + (\rho_m)_0 a_0^3 a(t)\right]\right]^{1/2}} \equiv \sqrt{\frac{\Lambda}{3}} \cdot \int dt \quad (41)$$

The following polynomial expression is obtained for scale factors near the big bang, assuming $u = a^{-1}$:

$$u^9 + A_1 \cdot \frac{u^8}{a_0} + A_2 \cdot \frac{u^7}{a_0^2} - A_3 \cdot \left(\frac{\Lambda}{8\pi}\right) \cdot \frac{u^5}{a_0^4} - A_4 \cdot \left(\frac{\Lambda}{8\pi}\right) \cdot \frac{u^4}{a_0^5} + A_5 \cdot \left(\frac{\Lambda}{8\pi}\right)^2 \cdot \frac{u^1}{a_0^8} + A_6 \cdot \left(\frac{\Lambda}{8\pi}\right)^2 \cdot \frac{t}{a_0^9} \cong 0 \quad (42)$$

One could calculate considerably higher polynomial roots of Eqn.(41) above, depending on the required degree of accuracy. This assumes a non-infinite value of $u = a^{-1}$, and non-infinite value for the $\Lambda$ term. This above equation uses the following coefficients.

$$A_1 = \frac{9}{4} \cdot \frac{(\rho_m)_0}{(\rho_{rel})_0}, \quad A_2 = \frac{(\rho_m)_0^2}{(\rho_{rel})_0^2}, \quad A_3 = \frac{1/5}{(\rho_{rel})_0^1}, \quad A_4 = \frac{(\rho_m)_0 \cdot (1/4)}{(\rho_{rel})_0^2}, \quad A_5 = \frac{1}{(\rho_{rel})_0^2}, \quad A_6 = \frac{1}{(\rho_{rel})_0^2} \cdot \sqrt{\frac{\Lambda}{3}} \quad (43)$$

It is argued that the existence of such a nonlinear equation for early universe scale factor evolution introduces a de facto "information" barrier between a prior universe (which as is argued here can only include thermal bounce input) and the new nucleation phase of our present universe. To see this, refer to Dowker (2005) on causal sets that require the following ordering, with a relation $\prec$, where it is assumed that initial relic space-time is replaced by an assembly of discrete elements to initially create a partially ordered set $C$

(1) If $x \prec y$, and $y \prec z$, then $x \prec z$

(2) If $x \prec y$, and $y \prec x$, then $x = y$ for $x, y \; \varepsilon \; C$

(3) For any pair of fixed elements $x$ and $z$ of elements in $C$, the set $\{y \mid x \prec y \prec z\}$ of elements lying in between x and z is finite

Statements (1) and (2) suggest that $C$ is a partially ordered set; statement (3) permits local finiteness. This when combined with as a model for how the universe evolves via a scale factor equation permits us to write, after we substitute $a(t^*) < l_P$ for $t^* < t_P =$ Planck time, and $a_0 \equiv l_P$, and $a_0/a(t^*) \equiv 10^\alpha$ for $\alpha \gg 0$ into a discrete equation model of Eqn (40) leads to

$$\left[\frac{a(t^* + \delta t)}{a(t^*)}\right] - 1 < \frac{(\delta t \cdot l_P)}{\sqrt{\Lambda/3}} \cdot \left[1 + \frac{8\pi}{\Lambda} \cdot \left[(\rho_{rel})_0 \cdot 10^{4\alpha} + (\rho_m)_0 \cdot 10^{3\alpha}\right]\right]^{1/2} \xrightarrow[\Lambda \to \infty]{} 0 \quad (44)$$

So in the initial phases of the big bang, with a very large vacuum energy, the following relation is obtained, which violates (signal) casuality. For any given fluctuation of time in the "positive" direction,

$$\left[\frac{a(t^* + \delta t)}{a(t^*)}\right] < 1 \quad (45)$$

It may be argued that the existence of a violation of causal set arrangements in the evolution of a scale factor argues for a discontinuity in the propagation of information from a prior universe to our present universe. This complements and adds to the importance of transfer of a large amount of thermal vacuum energy from a prior universe to relic inflationary conditions in today's universe. This paper argues that the existence of a wormhole for such thermal transferal of thermal energy from a prior universe to today's universe implies faster-than-light propagation. When such propagation of faster than the speed of light occurs, this argues for no information transfer between the initial and final starting points during the wormhole transfer of thermal energy. This in turn implyies causal discontinuity.

## CONCLUSIONS

This paper has tried to reconcile two models: (1) Brane world models will not permit an Ashtekar (2006) quantum bounce. This bounce idea is used to indicate how one can reconcile axion physics with the production of dark matter/dark energy later on in the evolution of the inflationary era, where one sees Guth style chaotic inflation for times $t \geq t_P$ and the emergence of dark energy during the inflation era. (2) The Carroll &Chen (2005) paper assumes a low entropy-low temperature pre-inflationary state of matter prior to the big bang. It addresses how to ramp up to the high energy values greater than $10^{12}$ Kelvin temperatures during nucleosynthesis. For suggested future research, this paper makes reference to solving the cosmological "constant" problem without using the Gurzadyan (2003) approach, which is fixed on the scale factor $a(t)$ for a present value of the cosmological constant.

One of the challenges ahead will be to link the onset of graviton production with the critical threshold energy, which is asserted to be a trigger for graviton production. In doing this, the authour will be looking, via the Sach-Wolfe effect, not only for evidence for the higher frequency range of Graviton radiation, but also for experimental evidence for the existence of short-term quintessence. Furthermore, Eqn (6) suggests a dramatically lowered value for the net range of graviton frequencies, if the initial volume of space for graviton production is localized in the regime near the Planck time interval. I.e., for information theory reasons, one needs to go out to the $z \approx 1100$ red shift limit years after the big bang to commence a region of space consistent with Eqn (6), which has high net graviton frequencies.. In other words, the Hartle-Hawking wave function comes into being right after graviton production, whereas it is zero beforehand. This would lead to a favored state for the nucleation of the cosmological landscape due to changes in the cosmological constant at or before a Planck's time interval, where we would have $z \sim 10^{26}$. In contrast, the region of space where $z \approx 1100$ would be required in order to be consistent with the Lloyd (2002) measurement of the computational space-time limits of how the universe evolves in time. Finally, this would lead to a theoretical understanding of how the scale factor would evolve initially. The chaotic conditions for scale factors as given in Eqn. (42) and Eqn (45) imply a causal relationship discontinuity between a prior universe and our present universe, as it is evolving A bifurcation can be observed between initial conditions of an exponential scale factor due to inflation, leading to a zero quintessence field, and another branch in the region where the scale factor, again has exponential expansion, due to a reduced cosmological constant. Signal propagation violation is implied by Eqn (45), which marks the delineation between the two forks of a scale factor bifurcation, indicating the causal breakdown in information between a prior universe, with thermal input into our present universe. A Wheeler-De Witt pseudo time dependent equation is also presented, with a solution dominated by a high temperature cosmological vacuum, delineating heat transfer between a prior and the present universe. Such a solution, due to the symmetry between present and past times, as given by Eqn. (24), is a symmetry, with the symmetry broken by relic graviton production. Symmetry-breaking also allows for examination of more of the quantum gravity geometry in early universe conditions. Furthermore, the causal breakage implied by Eqn (45) will be explored more thoroughly, with a view toward comparing it to pre-inflation conditions.